\documentclass[pre,aps,floatfix,reprint,onecolumn,superscriptaddress]{revtex4-1}
\usepackage{amsmath}
\usepackage{amssymb}
\usepackage{graphicx}
\usepackage{float}

\newcommand{\vp}{\varphi}
\newcommand{\w}{\omega}

\begin{document}
\title{Disentangling Respiratory Sinus Arrhythmia in Heart Rate Variability Records}
\author{\c{C}a\u{g}da\c{s} Top\c{c}u} 
\affiliation{Physiology, Otto Loewi Research Center for Vascular Biology,\\ 
 Immunology and Inflammation, Medical University of Graz,\\
  Neue Stiftingtalstr. 6/D05, A-8010 Graz, Austria}
\affiliation{Department of Physics and Astronomy, University of Potsdam,\\
 Karl-Liebknecht-Str. 24/25, D-14476 Potsdam-Golm, Germany}
\author{Matthias Fr\"uhwirth}
 \affiliation{Human Research Institute of Health Technology\\
 and Prevention Research,\\ Franz Pichler Street 30, A-8160 Weiz, Austria}
\author{Maximilian Moser}
\affiliation{Physiology, Otto Loewi Research Center for Vascular Biology,\\ 
 Immunology and Inflammation, Medical University of Graz,\\
  Neue Stiftingtalstr. 6/D05, A-8010 Graz, Austria}
 \affiliation{Human Research Institute of Health Technology\\
 and Prevention Research,\\ Franz Pichler Street 30, A-8160 Weiz, Austria}
\author{Michael Rosenblum} 
\author{Arkady Pikovsky}
\affiliation{Department of Physics and Astronomy, University of Potsdam,\\
 Karl-Liebknecht-Str. 24/25, D-14476 Potsdam-Golm, Germany}
\affiliation{The Research Institute of Supercomputing,\\ 
 Lobachevsky National Research State University\\ of Nizhny Novgorod, Russia}
\date{\today}
 
  \begin{abstract}
Different measures of heart rate variability and particularly of
respiratory sinus arrhythmia are widely used in research and clinical
applications. 
Inspired by the ideas from the theory of coupled oscillators, we use  
simultaneous measurements of respiratory and cardiac activity to perform
a nonlinear decomposition of the heart rate variability into the 
respiratory-related component and the rest. 
We suggest to exploit the technique as a universal preprocessing tool,
both for the analysis of respiratory influence on the heart rate 
as well as in cases when effects of other factors on the heart rate 
variability are in focus.
The theoretical consideration is illustrated by the analysis of 25 data 
sets from healthy subjects.
  \end{abstract}
 \maketitle
  
\section{Introduction}
Heart rate variability (HRV) is a non-invasive measure of autonomic nervous system function.
Therefore, its analysis and quantification are increasingly 
used in physiological and medical research as well as in clinical practice.
Typically, HRV denotes variation of the inter-beat intervals, in most cases 
defined as the intervals between the well-pronounced R-peaks in an  
electrocardiogram (ECG), and therefore called RR-intervals.
A particularly important component of HRV is a modulation of the RR-intervals
by respiratory influence, called respiratory sinus arrhythmia 
(RSA)~\cite{Katona-Jih-75,Hirsch-Bishop-81,Hayano842,Billman-11,Moser1994}.
Physiological significance of RSA is, on one hand, in facilitating 
gas exchange between the lungs and the blood and thus 
helping the heart to do less work while maintaining optimal levels of blood 
gases~\cite{Larsen_et_al-10,TJP:TJP4976}.
On the other hand, biological advantage of RSA is the stabilization of blood flow 
to the brain and the periphery by compensating  arterial pressure changes 
arising from intrathoracic pressure changes due to in- and expiration. 
It has been shown that blood pressure oscillations, connected to respiration, 
are reduced and, therefore, the blood flow is stabilized by 
RSA~\cite{Elstad2014}. 
In medicine, the amplitude or the spectral power of RSA is used as a smart noninvasive 
measure for vagal tone~\cite{Moser1994}. 
The reason for this dominantly vagal origin of RSA can be found 
in the vagal synapses, which are faster than the sympathetic ones and are therefore able 
to translate central respiratory oscillations present in the brainstem 
to changes of cardiac sinus node discharge 
rate, which is not the case for slow sympathetic synapses~\cite{Moser2017}. 

Vagal tone is gaining importance in preventive and aging medicine as the tone 
decreases with age~\cite{Baylis2013,Lehofer1999} and also because of chronic 
diseases~\cite{Lehofer1997,Das2001}. 
This understanding got momentum, when a vagal inflammatory reflex was 
discovered~\cite{Tracey2002}, 
indicating a close inverse connection between the available vagal tone and 
silent inflammation, a condition obviously resulting in chronic diseases like 
vascular sclerosis, Alzheimer disease, and even cancer (see~\cite{Nathan2002} 
for a comprehensive overview). Since the accuracy of vagal tone determination
by common time or frequency domain 
methods of RSA quantification, especially under conditions of different respiratory 
patterns, have been questioned~\cite{Laborde2017}, 
it is highly important to improve the methods for separation of respiratory 
and other influences of the autonomic nervous system on HRV. 
   
A variety of data analysis techniques quantifying RSA have 
been proposed in the literature; for a discussion of commonly used 
metrics and their advantages and drawbacks see, e.g.,~\cite{Lewis_et_al-12}. 
Examples of application of RSA analysis include clinical 
psychology~\cite{Wielgus-Aldrich-Mezulis-Crowell-16},
treatment of substance use disorder~\cite{Price-Crowell-16},
prediction of the course of depression~\cite{Panaite_et_al-16},
quantification of cardiac vagal tone and its relation 
to evolutionary and behavioral functions~\cite{Grossman-Taylor-07},
quantification of vagal activity during stress in 
infants~\cite{Ritz_et_al-12} and even in cancer patients~\cite{Moser2006}, to name just a few.
On the other hand, quantification of the HRV component, not  
related to respiration,   
is important for the analysis of long-range and scaling properties of the
cardiac dynamics~\cite{Ivanov_et_al_99a,Schmitt-Ivanov-07}.

In this paper we, following our previous study~\cite{Kralemann_et_al-13}, 
elaborate on a nonlinear technique that allows us to decompose the HRV 
into a respiratory-related component (R-HRV), and a component where variability is 
caused by all other sources; we denote the latter component as NR-HRV. 
After the decomposition, both components can be subject to any existing
analysis techniques. Thus, the suggested disentanglement can serve as a
universal preprocessing tool that allows a researcher to concentrate on 
particular aspects of HRV: if the interest is in a respiratory caused 
modulation of the heart rate, then it makes sense to work with the  
R-HRV component. 
On the contrary, if the variation of the heart rhythm in a different frequency 
range is important, then it makes sense to first cleanse the data from the
respiratory caused variability, and then to analyze NR-HRV.

Separation of respiratory influences from HRV records was also treated before by 
different \textit{ad hoc} techniques 
\cite{Widjaja_et_al-14,Kuo-Kuo-16} like adaptive filtering, least-mean-error
fitting of power spectra, and principal component analysis.
Our approach is based on the idea from nonlinear dynamics and coupled 
oscillators 
theory~\cite{Winfree-80,Glass-01,Pikovsky-Rosenblum-Kurths-01,Strogatz-03}. 
Within this framework, we treat cardio-vascular and 
respiratory systems as two interacting endogenous, self-sustained, oscillators, 
what allows for a low-dimensional description of their dynamics in terms of phases.
This description, in its turn, provides the desired  disentanglement and better
quantification of the corresponding HRV components, as described below.

\section{Methods}
\subsection{Theoretical Background}
We start with a general theory, briefly presenting  a phase-based description 
of the dynamics of interacting oscillators.
In the simplest case of an autonomous noisy periodic 
or weakly chaotic oscillator the phase dynamics obeys
\begin{equation}
\dot\vp=\w+\zeta(t)\;,
\label{eq:gpd0}
\end{equation}
where $\vp$ and $\w$ are the phase and the natural frequency of the system, and 
the noise term $\zeta(t)$ accounts for intrinsic fluctuations of the system parameters.
If the system experiences external influences from different sources (which may be 
either regular or not),
then, according to the dynamical perturbation theory
(see, e.g., \cite{Pikovsky-Rosenblum-Kurths-01} for details), 
the leading effect of the external forces $\eta_s(t)$, where $s=1,2,\ldots$, 
is in the modulation of the phase, which now obeys 
\begin{equation}
\dot\vp=\w+\sum_s q_s[\vp,\eta_s(t)]+\zeta(t)\;.
\label{eq:gpd}
\end{equation}
Here $q_s$ are coupling functions
describing response to the corresponding perturbations.
It accounts for the property that susceptibility of an oscillator to external 
perturbations generally depends on its phase.
Notice that the forces $\eta_s(t)$ can have arbitrary complex time dependence, 
i.e. they can be periodic, chaotic, or stochastic. 

Suppose now that one of the forces, say the first one, $\eta_1(t)$, and the 
corresponding coupling function $q_1$, are known.
Then we can use Eq.~(\ref{eq:gpd}) in order to represent the variations of the 
instantaneous frequency as a sum of two components,
\begin{equation}
\dot\vp=\w+q_1(\vp,\eta_1(t))+\xi(\vp,t)\;,
\label{eq:gpd1}
\end{equation}
where 
$
q_1(\vp,\eta_1(t))
$
describes  solely the impact of the 
force $\eta_1(t)$ on the phase dynamics, while 
\begin{equation}
\xi(\vp,t)= \sum_{s\ne 1} q_s(\vp,\eta_s(t))+\zeta(t)
\label{eq:gpd3}
\end{equation}
describes a cumulative effect of all other forces and of the intrinsic fluctuations.
The representation by Eq.~(\ref{eq:gpd1}) plays the central role in our approach.

Now we specify the theory to cover the system of our interest, namely the
cardiovascular system.
In particular, we consider the experiments where both 
cardiac and respiratory activities are monitored simultaneously, and the
ECG and the respiratory signal, e.g. the air flow at the nose,
are registered. It is natural to represent these two endogenous rhythms as
outputs of two interacting oscillatory systems, which can be characterized 
by their phases.
As discussed in details below, these phases can be estimated from data.

Denoting the phase and the natural frequency of the cardiac 
oscillator with $\vp$ and $\w$, respectively, we write, similarly to \eqref{eq:gpd1}:
\begin{equation}
\dot\vp=\w+q_r(\vp,\eta_r(t)) + \xi(\vp,t)\;,
\label{eq:gpd4}
\end{equation}
where the subscript $r$ stands for respiration, and $\eta_r$ describes effect of the 
respiration on the cardiac frequency. The term $\xi(\vp,t)$, like in Eq.~(\ref{eq:gpd3}),
describes the effect on the cardiac phase of the physiological rhythms other than 
respiration, as well as of non-rhythmical, stochastic, external and intrinsic perturbations.

In practice, measurements of respiration rather often (and in our experiments as well)
do not provide a proper magnitude 
of the corresponding force $\eta_r$, but only its phase $\psi(t)$. 
Thus, we assume that the forcing term due to respiration is a $2\pi$-periodic function
of the time-dependent respiratory phase $\psi$. So, we write 
$\eta_r(t)=S(\psi(t))=S(\psi(t)+2\pi)$,
and replace the coupling function 
$q_r(\vp,\eta_r(t))$ in Eq.~(\ref{eq:gpd4}) by 
a phase-based coupling function 
$Q(\varphi,\psi)$, obtaining thus
\begin{equation}
\dot\vp=\w+ Q(\varphi,\psi)+ \xi(\vp,t)\;.
\label{eq:gpd7}
\end{equation}

In order to introduce the main idea of our paper, we postpone the discussion of how 
the terms in Eq.~(\ref{eq:gpd7}) can be obtained from data, and assume for the moment 
that they are already known. 
Next, we focus on a link between the phase dynamics description via Eq.~\eqref{eq:gpd7} 
and the standard representation of the HRV via a sequence of the RR-intervals.
We emphasize that the phase $\vp$ can be always introduced in such a way, that 
\begin{equation}
\vp(t_k)=2\pi k,
\label{eq:gpd8}
\end{equation}
where $t_k$ is the instant of appearance of the $k$-th R-peak. 
If some other definition of the phase is used, addition of 
a constant phase shift ensures property~(\ref{eq:gpd8}). 
Notice also that the phase can be in an equivalent 
way considered as a variable wrapped to $[0,2\pi)$ interval; in this representation 
the phase achieves the value $2\pi$ and immediately resets to zero at the instant 
of an R-peak appearance. 
Thus, knowledge of the phase evolution $\vp(t)$ yields RR-intervals and hence
fully determines HRV.

Exploiting the representation \eqref{eq:gpd7}, we now introduce two new phases. The first
one, $\Phi$, describes  solely the effect of the respiration on the instantaneous 
cardiac frequency, and obeys
\begin{equation}
\dot\Phi=\w+Q(\Phi,\psi)\;.
\label{eq:prsa}
\end{equation}
This equation is obtained from Eq.~(\ref{eq:gpd7}) by dropping the last term.
Correspondingly, the other phase, $\Psi$, describes the effect of all 
other forces, except for respiration, and of internal fluctuations, on the heart rate. 
This phase is governed by
\begin{equation}
\dot\Psi=\w+\xi(\Psi,t)\;.
\label{eq:pnrsa}
\end{equation}
In fact, because the time series $\vp(t),\psi(t),\xi(\vp,t)$ are known from the 
processing of measured data, Eqs.~(\ref{eq:prsa},\ref{eq:pnrsa}) can be
straightforwardly integrated to yield time series $\Phi(t)$ and $\Psi(t)$. 
(Practically, we used the Euler integration scheme with initial conditions
$\Phi(t_1)=0$, $\Psi(t_1)=0$.)

\begin{figure}[!hbt]
\centering
\includegraphics[width=0.95\textwidth]{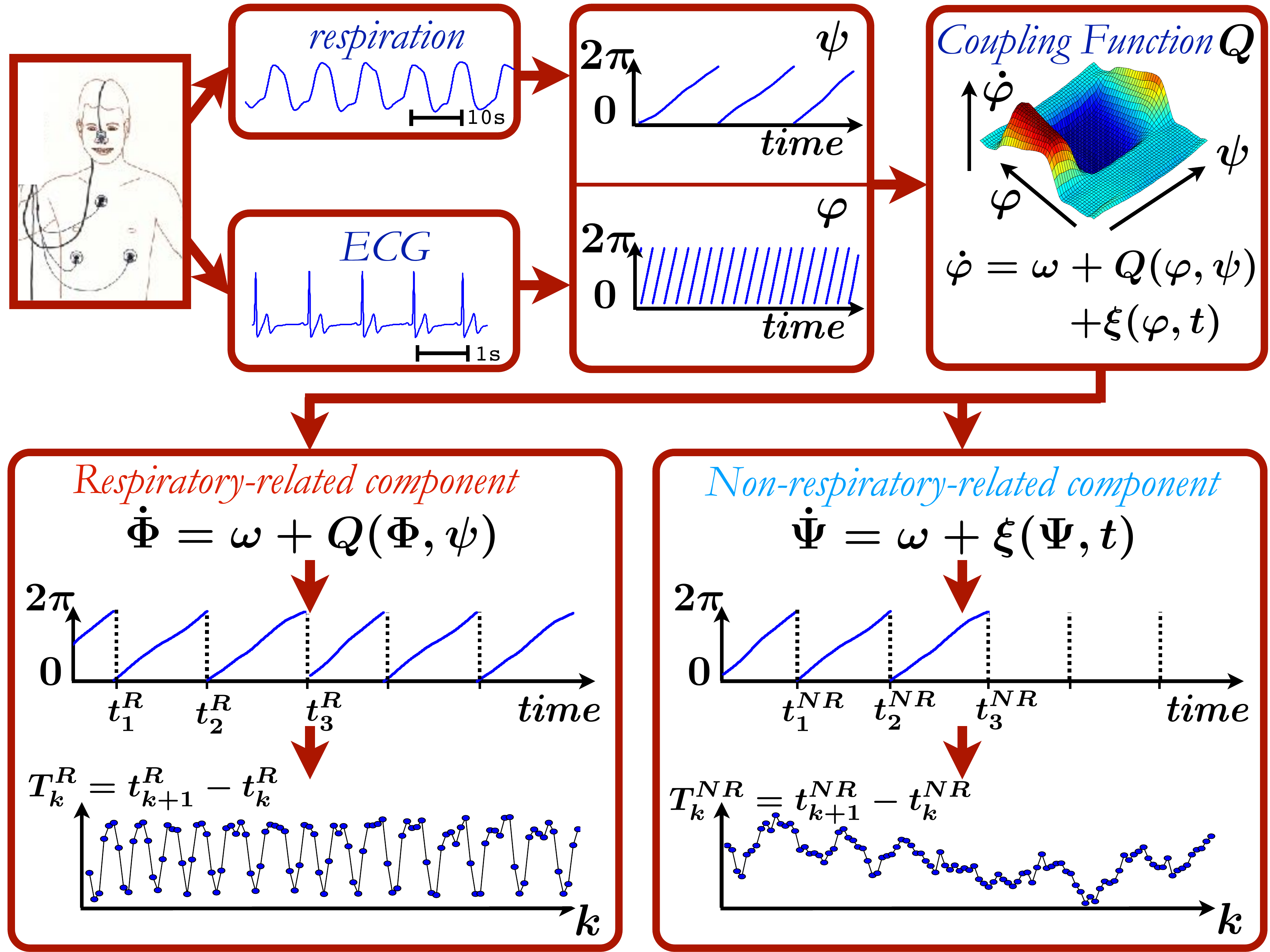}
\caption{The approach at a glance. 
In the first step, instantaneous phases, 
$\vp$ and $\psi$, are obtained from simultaneously measured ECG and respiratory signals. 
The cardiac phase $\vp$ is defined in such a way that values 
$\vp=2\pi k$ correspond to the R-peaks (wrapped definition of the phase is used here 
for better visibility). 
Next, the phase dynamics 
is reconstructed from data in the form of Eq.~(\ref{eq:gpd7}) with the coupling 
function $Q$ and the rest term $\xi$. 
This equation yields equations for instantaneous frequency of respiratory-related 
and non-respiratory-related components, i.e. for $\dot\Phi$ and $\dot\Psi$. 
Numerical integration of the latter equations provides phases of these components,
while the conditions $\Phi(t_k^R)=2\pi$ and $\Psi(t_k^{NR})=2\pi$ determine the 
newly generated series of R-peaks. As a final result we obtain tachograms for 
both components. 
}.
\label{scheme}
\end{figure}

Knowing the phases $\Phi,\Psi$ one can easily obtain the RR intervals, using 
definition \eqref{eq:gpd8}.
The times at which the phase $\Phi$ attains a value that is a multiple of $2\pi$, i.e. when 
$\Phi(t_k^{R})=2\pi k$, yield a series of R-peaks, as it would look like in 
the presence of respiratory influence only. 
Thus, this series $t_k^{R}$ and the corresponding series of 
RR-intervals $t_{k+1}^{R}-t_k^{R}$ represent the pure RSA-related component, R-HRV,
 of HRV. 
On the contrary, if we are interested in the HRV due to all sources except for 
RSA, then we use the phase $\Psi$
to obtain the series of R-peaks, determined by the instants $t_k^{NR}$,
such that $\Psi(t_k^{NR})=2\pi k$, 
and the RR-intervals which vary due to non-respiratory influences only.
This completes disentanglement of the HRV into the R-HRV component 
(series of R-peaks at $t_k^{R}$ and the corresponding tachogram  
$T_k^{R}=t_{k+1}^{R}-t_k^{R}$)
and into the
NR-HRV component (series of R-peaks at $t_k^{NR}$ and
the corresponding tachogram  
$T_k^{NR}=t_{k+1}^{NR}-t_k^{NR}$).
Schematic illustration of the approach is given in Fig.~\ref{scheme}.

\subsection{Reconstruction of the phase dynamics from data}
We analyze below the set of 25 records of $420$ seconds long simultaneous 
measurements of respiratory flow  and ECG, already explored for different purposes 
in the previous publication~\cite{Kralemann_et_al-13}.
The experiments were performed on healthy adults at rest, in a supine position. 
The details of the measurements, the preprocessing, and the subjects are 
described in the Supplementary Information to this paper 
and to Ref.~\cite{Kralemann_et_al-13}.

Now we describe the particular steps
behind the general disentanglement approach presented above. Since
the representation of the cardiac dynamics via Eq.~\eqref{eq:gpd7} 
has been derived
in~\cite{Kralemann_et_al-13}, 
here we
only briefly outline the main steps.
\begin{enumerate}
\item Recorded cardiac ECG signals and respiratory signals have been embedded
in a two-dimensional plane by virtue of the Hilbert transform. 
The protophase (phase-like variable) of the
respiratory signal has been obtained as an angle in this plane. The protophase
of the ECG signal required an extended processing, because this signal has 
a complex form with several loops over a basic cycle. In fact, in these approaches
the details of parameterisation which may influence the definition of a protophase
are not important, as is explained in the next lines.
\item A transformation from the protophases to the phases has been performed, according
to the method suggested in Ref.~\cite{Kralemann-08}. The main idea is that because
the embedding and the parameterisation of the embedded trajectory by a $2\pi$-periodic
phase-like variable are not unique, the obtained protophase is not unique too. The true
phase is determined to have the property of growing linearly in time in absence of 
external forces (cf. Eq.~\eqref{eq:gpd0} which is written for the true phase and 
thus the deterministic part of the r.h.s.
does not depend on $\vp$). We have performed an invertible 
deterministic transformation
from the protophase to the phase (Eq. (4) in Ref.~\cite{Kralemann_et_al-13}), 
based on the property that the probability distribution
density of the phase should be uniform.
\item Having now the time series of the true phases of the cardiac system, $\vp(t)$, 
and of the  respiratory one, $\psi(t)$, we have calculated the time derivative 
$\dot\vp$ and have fitted it according to Eq.~\eqref{eq:gpd7} with a function, 
which is $2\pi$-periodic in arguments $\vp,\psi$. 
Practically, a kernel estimation (Eq. (7) in 
Ref.~\cite{Kralemann_et_al-13}) has been employed. 
As a result of this step,
the basic Eq.~\eqref{eq:gpd7} describing the cardiac phase dynamics is reconstructed.  
\end{enumerate}

\subsection{Characterization of original and disentangled HRV data}
In order to quantify the original RR-intervals and the results of the 
disentanglement procedure, 
we computed for all 25 data sets several physiologically relevant  measures 
that are commonly used in HRV analysis.

The statistical measure in time-domain that directly characterizes 
the ``evenness'' of the sequence of RR intervals,
is RMSSD: root mean square of successive differences \cite{Guidelines1996}, 
defined as
\[
\text{RMSSD}=\sqrt{\langle|T_{k+1}-T_k|^2\rangle}\;.
\]
Another measure, LogRSA~\cite{Lehofer1997}, is defined as logarithm of the median 
of the distribution of the absolute values of successive differences, i.e. 
\[
\text{LogRSA}=\log \left[ \text{median}|T_{k+1}-T_k|\right]
\]
LogRSA takes care for a nearly log-normal statistical distribution of the medians, so
that by taking logarithm a nearly normal distribution is achieved.
Several studies~\cite{Lehofer1999,Lehofer1997,Grote2007,Moser1998} have proven 
robustness of LogRSA and its ability to differentiate between 
vagal states. 
Another characteristics of ``non-evenness''
of RR interval series used especially in clinical settings is the relative 
(i.e. divided by the total number of the RR 
intervals in the time series)
number of successive pairs of RR-intervals, that differ by more than 50 ms, denoted 
as  $\text{pNN50}$ \cite{Guidelines1996}. 
Finally, we compute the standard deviation of RR intervals, SDNN. 

In the frequency domain methods based on the power spectral density of 
the time series of RR intervals, 
one commonly computes the power in three frequency
bands: VLF (very low frequency, from 0.0033 to 0.04 Hz), LF 
(low frequency, from 0.04 to 0.15 Hz), and 
HF (high frequency, from 0.15 to 0.4 Hz), e.g. by Fourier analysis. 

Moreover, many nonlinear measures, mainly based on  dynamical system 
approaches, have been applied to characterize HRV. 
Some of these measures require rather long time series and are 
therefore not applicable to our relatively short observations.
As appropriate indices we have calculated the approximate entropy 
(ApEn)~\cite{Pincus1991a,Pincus1991b} 
and the sample entropy (SampEn)~\cite{Richman-Moorman-00,Chen_etal-09} 
of the HRV time series. (The tolerance value was taken as 15\% of the standard deviation
and the embedded dimension was fixed at 2.)

\section{Results}

\subsection{Time series and spectra}
\begin{figure}[!hbt]
\centering
\includegraphics[width=0.9\textwidth]{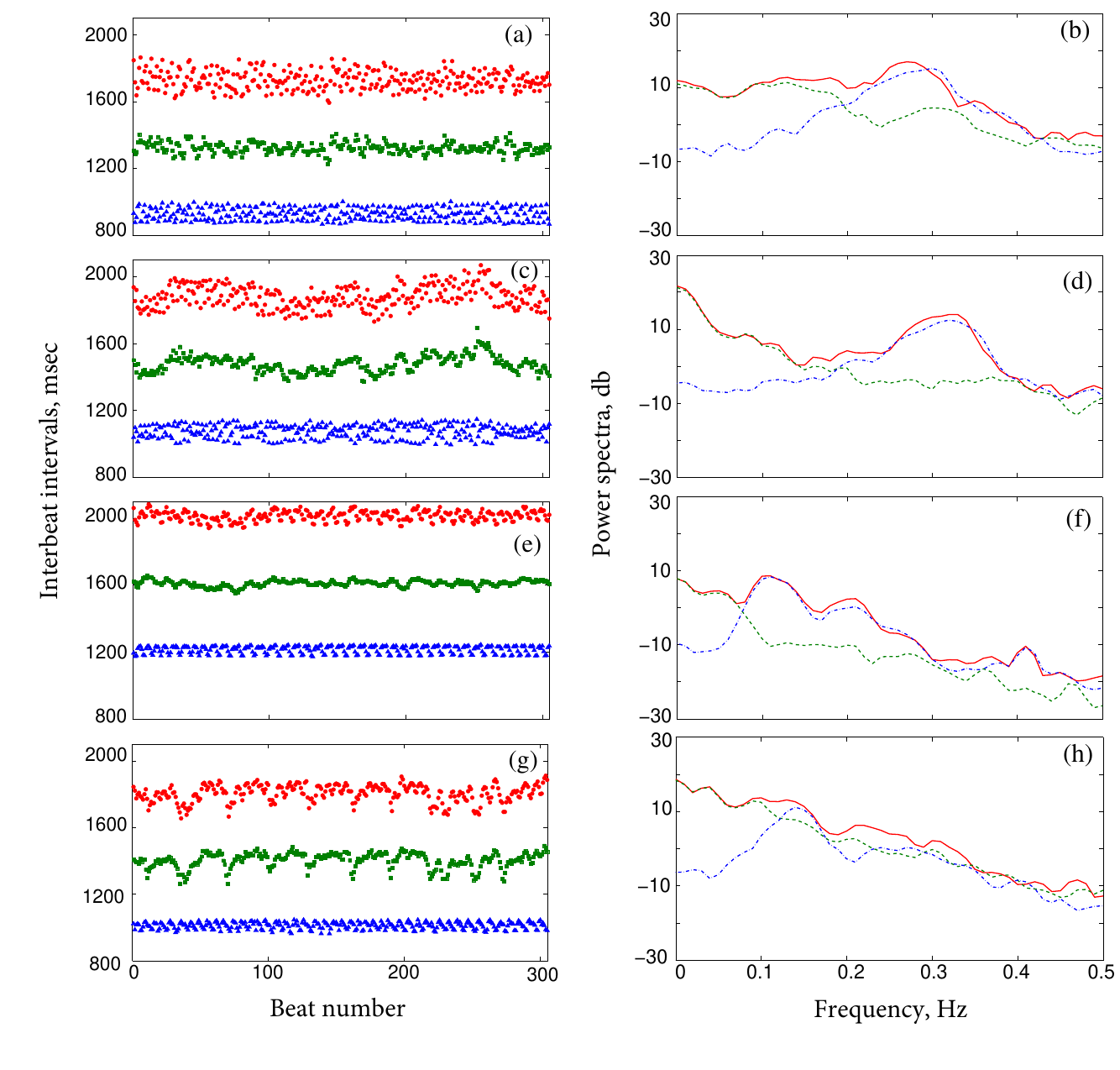}
\caption{HRV plots (left column) and corresponding power spectral densities (right column).
Original tachogram (red circles) is shown along with its R-HRV (blue triangles) 
and NR-HRV components (green squares). The corresponding spectra are given by 
red solid, blue dot-dashed, and green dashes lines, respectively.
Left, for better visibility, RR intervals for NR-HRV component 
and for the original time series
are shifted vertically by 400 ms and 800 ms, respectively.
For the same reason only 300 beats are shown in tachograms; full records are 
shown in Supplementary Material. For discussion, see text.}
\label{fig:rrintsp}
\end{figure}

First, we illustrate the method, presenting the original HRV series $T_k$
along with the disentangled components $T_k^{R},T_k^{NR}$
in Fig.~\ref{fig:rrintsp}(a,c,e,g).  
In each panel these three series
are shown by different colors and markers, and additionally shifted vertically 
for better visibility. We have chosen for presentation these four 
characteristic cases, while all studied cases
are presented in the Supplementary Material. 
In all the cases, the original HRV time series show different extent of 
interval-to-interval variability. 
The R-HRV time series also show significant variability; these time series 
are however much more homogeneous just by their construction: 
the term $Q$ in Eq.~\eqref{eq:prsa} has a definite constant
amplitude which is reflected in the magnitude of the variations of the 
R-HRV component. However, one can see clear differences in the NR-HRV
series, obtained via Eq.~\eqref{eq:pnrsa}.
Now we go beyond visual inspection and quantify the components.

Basically, we can distinguish two types
of the NR-HRV data: ``smooth'' and ``rough''. 
In the former case, the differences between
successive RR intervals are small, and the whole graph looks like a curve, possibly
slightly perturbed. To this case belongs Fig.~\ref{fig:rrintsp}(e). 
In the ``rough'' case, the differences between the subsequent intervals are 
large, and one does not see a curve, but rather a 
dispersed set of points (panel (a)). There are also intermediate cases, 
like in panel (c). Finally, in panel (g) we show a remarkable NR-HRV pattern, 
consisting of ``smooth''
patches interrupted nearly periodically (approx. at every 30-th heart beat) 
by ``rough'' bursts.

A complementary information is contained in the power spectra of the HRV
records, computed according to the procedure described in \cite{Clifford2005}  
and shown in the right column of Fig.~\ref{fig:rrintsp}. 
One can clearly recognize several characteristic features:
\begin{enumerate} 
\item The respiratory-related R-HRV component has pronounced peaks. These peaks 
can be interpreted as the main 
frequency of respiration, its harmonics and the combinational 
harmonics, i.e. heartbeat frequency $\pm$
respiration frequency.
\item The NR-HRV part contains no pronounced peaks. This confirms that
the main \textit{regular} oscillatory contribution to the HRV is
the respiration; all other 
perturbations are rather noisy and do not exhibit noticable spectral peaks 
(for the case 
depicted in panels (g,h), the low-frequency periodicity cannot be resolved
in the power spectrum for such a short time series).
\item The R-HRV component has low values at frequencies smaller than that
of respiration. In fact, at these frequencies, the spectra of the original HRV and of the 
NR-HRV practically coincide. 
This means that the slow irregular variability of the heart rhythm
is mainly caused not by respiration, but by other physiological processes. Our 
disentanglement allows one to study these slow components in a reliable way, by
cleansing the respiratory component that may hide the interesting 
slow processes.
\end{enumerate}

\begin{figure}[!hbt]
\centering
\includegraphics[width=0.7\textwidth]{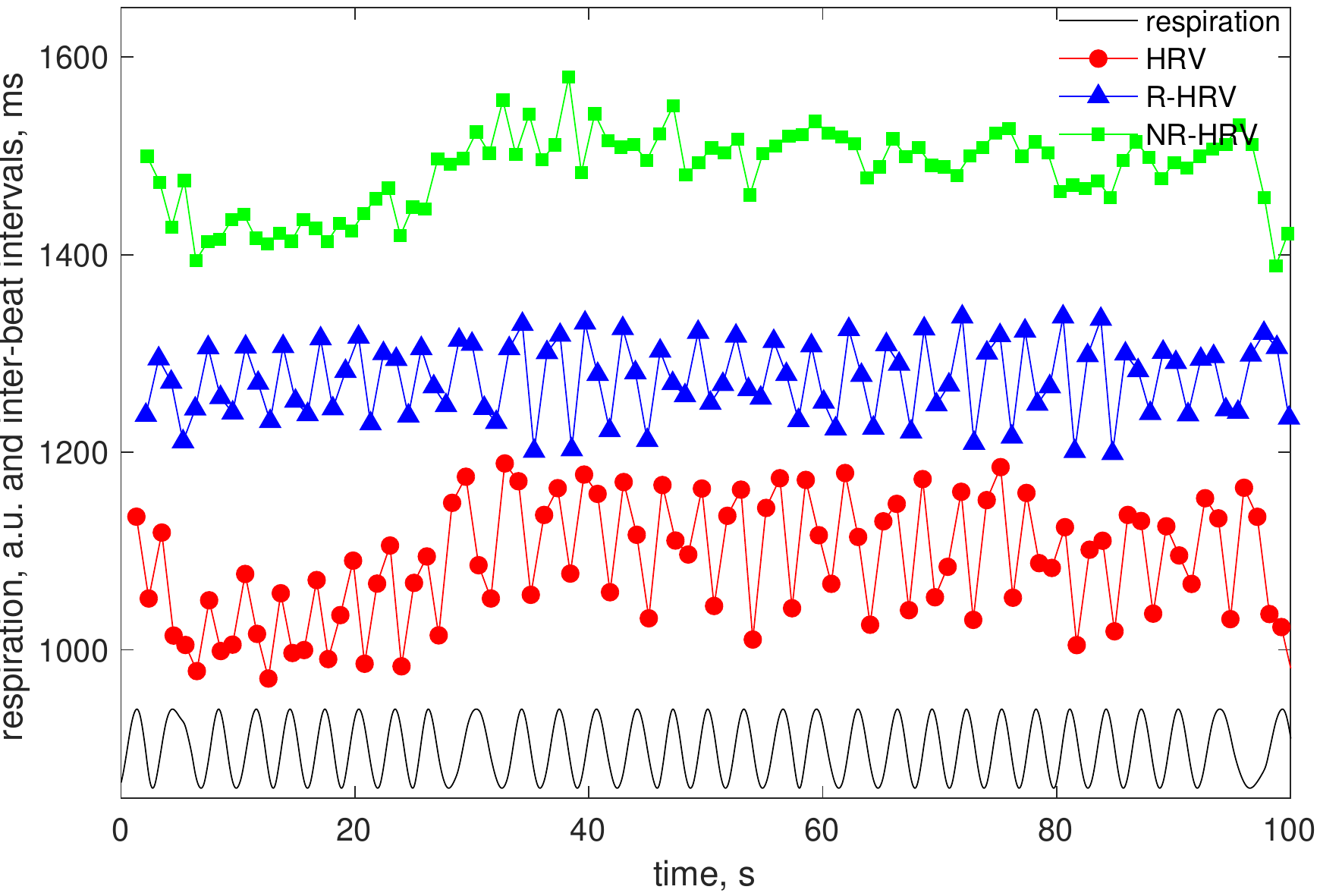}
\caption{Example of the original HRV and its respiratory and non-respiratory
related components. The latter two are shifted upwards for better visibility 
by 200 ms and 400 ms, respectively. 
Black curve shows respiratory signal with the amplitude 
normalized to one. 
}
\label{fig:examphrv}
\end{figure}

For another illustration of the disentanglement we plot in Fig.~\ref{fig:examphrv}
the original series and two constructed components components vs time, 
i.e. $T_k=t_{k+1}-t_k$ vs $t_{k+1}$, 
$T_k^R=t_{k+1}^R-t_k^R$ vs $t_{k+1}^R$, 
and $T_k^{NR}=t_{k+1}^{NR}-t_k$ vs $t_{k+1}^{NR}$, for one data set. 
Here we also show the time course of respiration, plotting $\cos\psi$ vs time.
One can see that though the R-peaks in the HRV (filled circles) and in R-HRV
(triangles) series occur at different instants of time,   
the overall pattern of the modulation by respiration, i.e. positions of peaks 
and  troughs in the tachograms, is preserved.   

\subsection{Time-domain characterizations of HRV}

\begin{figure}[!hbt]
\centering
\includegraphics[width=0.45\textwidth]{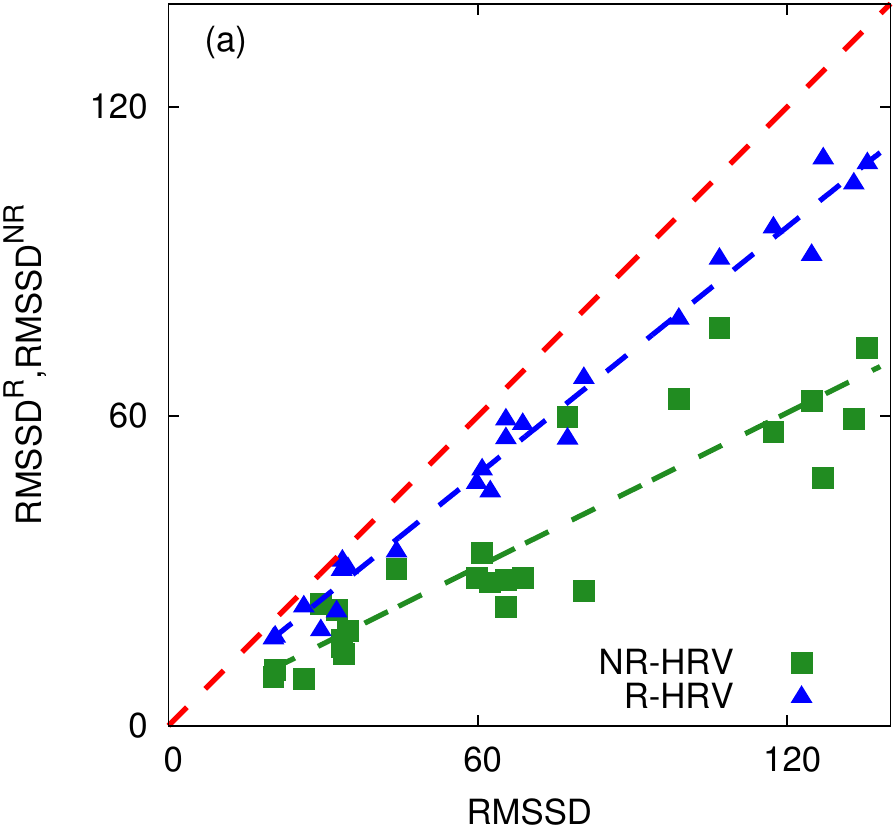}\hfill
\includegraphics[width=0.45\textwidth]{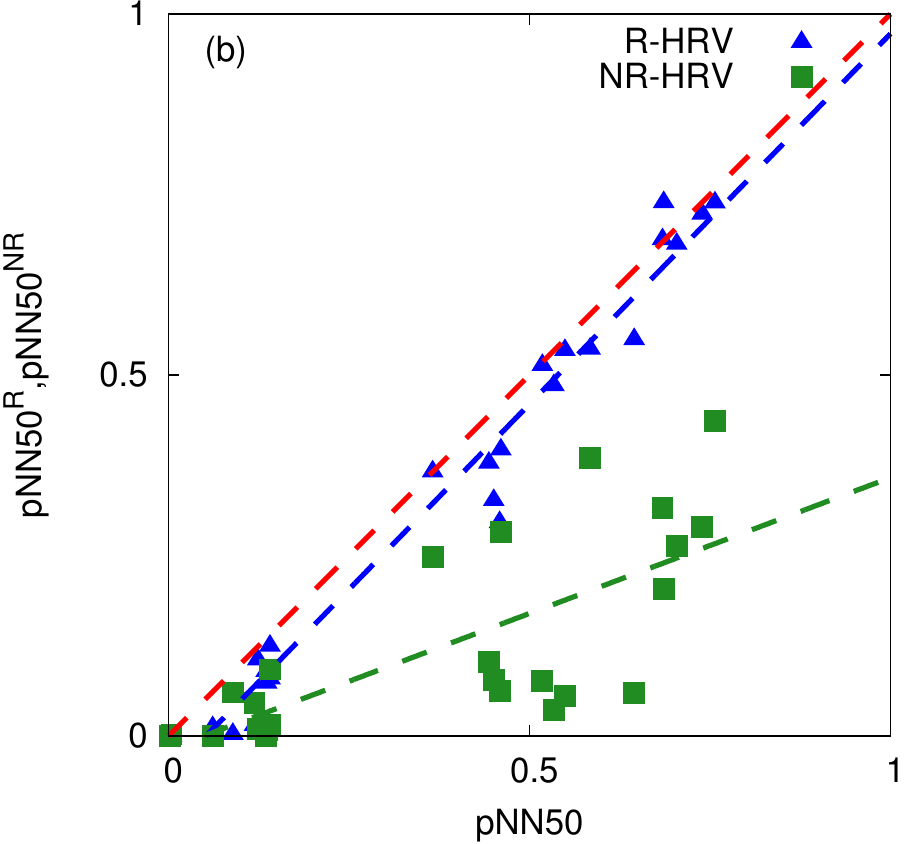}\\
~~\includegraphics[width=0.44\textwidth]{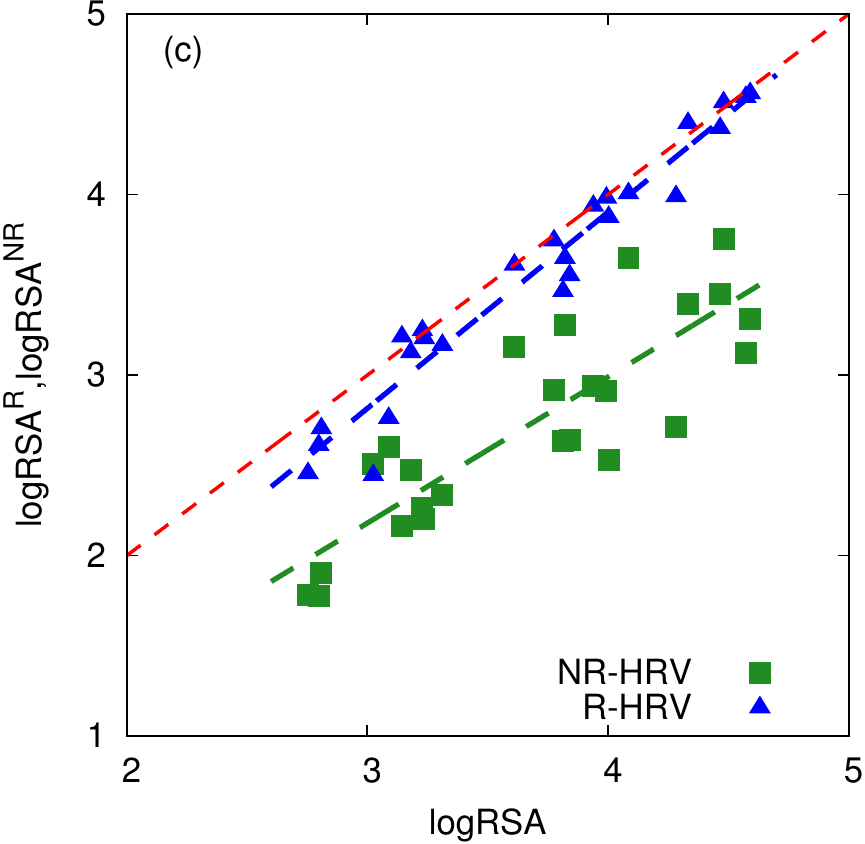}\hfill
\includegraphics[width=0.45\textwidth]{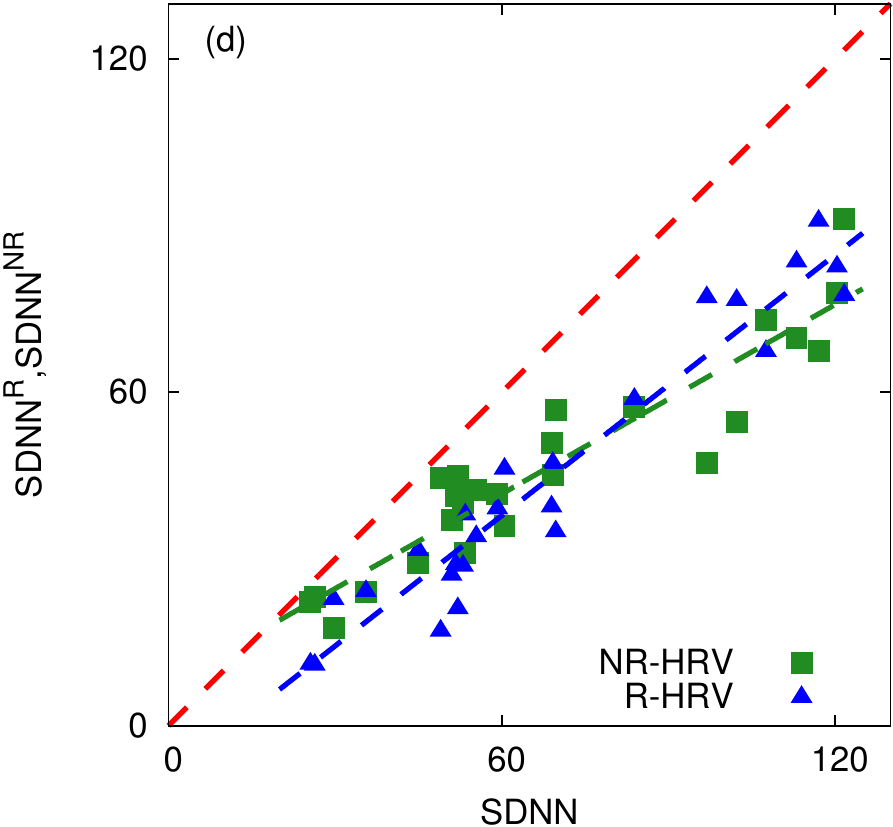}
\caption{(a): The values $\text{RMSSD}^{R}$ and  
$\text{RMSSD}^{NR}$ versus $\text{RMSSD}$. 
(b): The  values
of $\text{pNN50}^{NR}$ and $\text{pNN50}^{R}$
vs $\text{pNN50}$. 
(c): The  values of $\text{logRSA}^{NR}$ and $\text{logRSA}^{R}$
vs $\text{logRSA}$. 
(d): The  values of $\text{SDNN}^{NR}$ and $\text{SDNN}^{R}$
vs $\text{SDNN}$. 
For clarity, the line of identity is shown by the red dashed line. 
Other dashed lines show linear fits (the slopes and corresponding
correlation coefficients are given in square brackets): 
$\text{RMSSD}^{\text{NR-HRV}}$: $[0.50, 0.90]$;
$\text{RMSSD}^{\text{R-HRV}}$: $[0.90,  0.99]$;
$\text{pNN50}^{\text{NR-HRV}}$: $[0.38, 0.72]$;
$\text{pNN50}^{\text{R-HRV}}$: $[1.03, 0.99]$;
$\text{logRSA}^{\text{NR-HRV}}$: $[0.81, 0.85]$;
$\text{logRSA}^{\text{R-HRV}}$: $[1.09, 0.97]$;
$\text{SDNN}^{\text{NR-HRV}}$: $[0.57, 0.93]$;
$\text{SDNN}^{\text{R-HRV}}$: $[0.78, 0.96]$.
}
\label{fig:rmssd}
\end{figure}

Figure~\ref{fig:rmssd}(a) shows the RMSSD measure
for both disentangled components of all data sets.
We see that the RMSSD values for both the respiratory $\text{RMSSD}^{R}$ 
and non-respiratory $\text{RMSSD}^{NR}$ components are smaller than for 
the original one. 
We have found that in almost all cases, the reduction for the non-respiratory component
is much stronger than for the respiratory one.
One can see that the relative reduction of the RMSSD both for the R-HRV component (on 
average by factor 0.8) and NR-HRV component (on average by factor 0.5)
does not depend on the RMSSD value for the original HRV series. 
Panel (b) of the same figure presents the pNN50 measure of  ``non-smoothness''
of RR interval series. One can see that this quantity is significantly reduced
for the NR-HRV component, while its value for the R-HRV component is nearly the 
same as for original data. 
(Exception are cases where  pNN50 in the original time series is very small, here
the successive pairs with RR intervals larger than 50 ms are nearly eliminated both in 
the R-HRV and NR-HRV components). Panel (c) if Fig.~\ref{fig:rmssd} shows the values
of the measure logRSA. Reduction of this measure in the NR-HRV component is even more
evident than in panels (a,b). Finally, panel (d) demonstrates that 
both components of the $\text{SDNN}$ measure are reduced.

\subsection{Frequency-domain characterizations of HRV}

\begin{figure}[!hbt]
\centering
\includegraphics[width=0.48\textwidth]{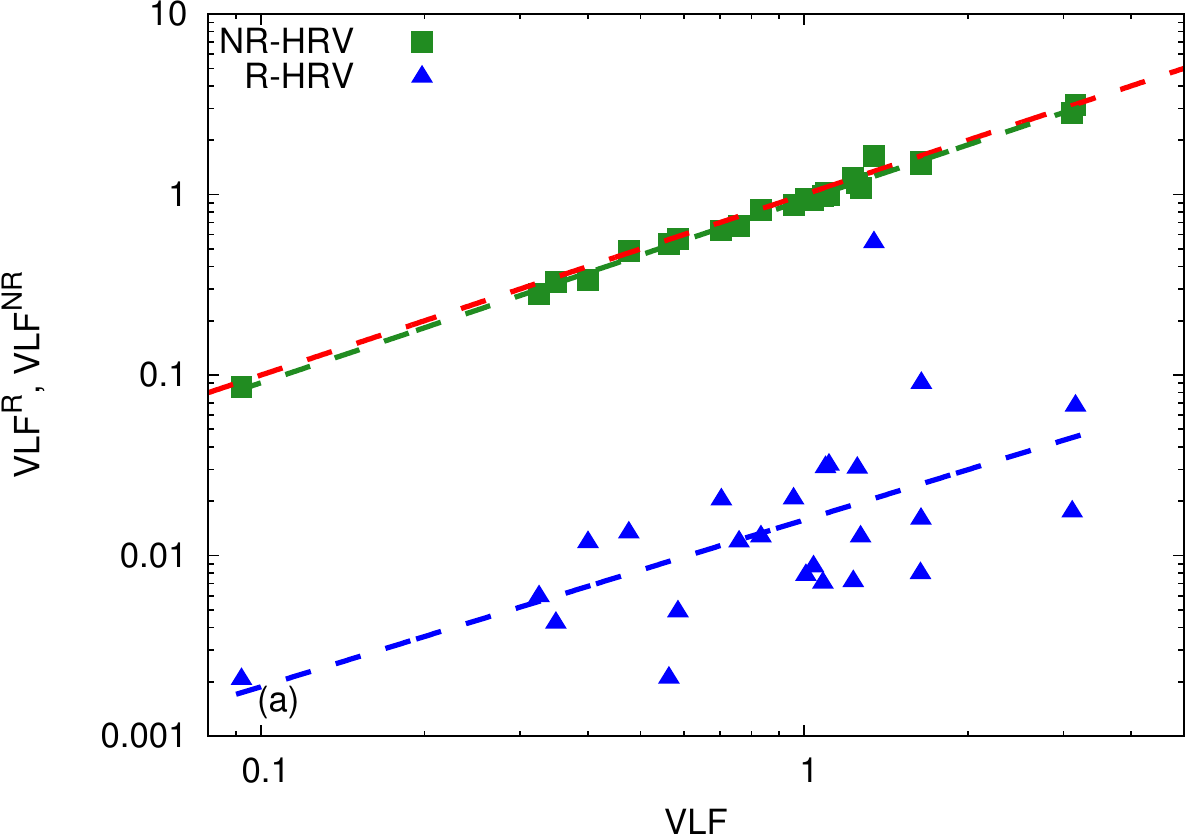}
\includegraphics[width=0.48\textwidth]{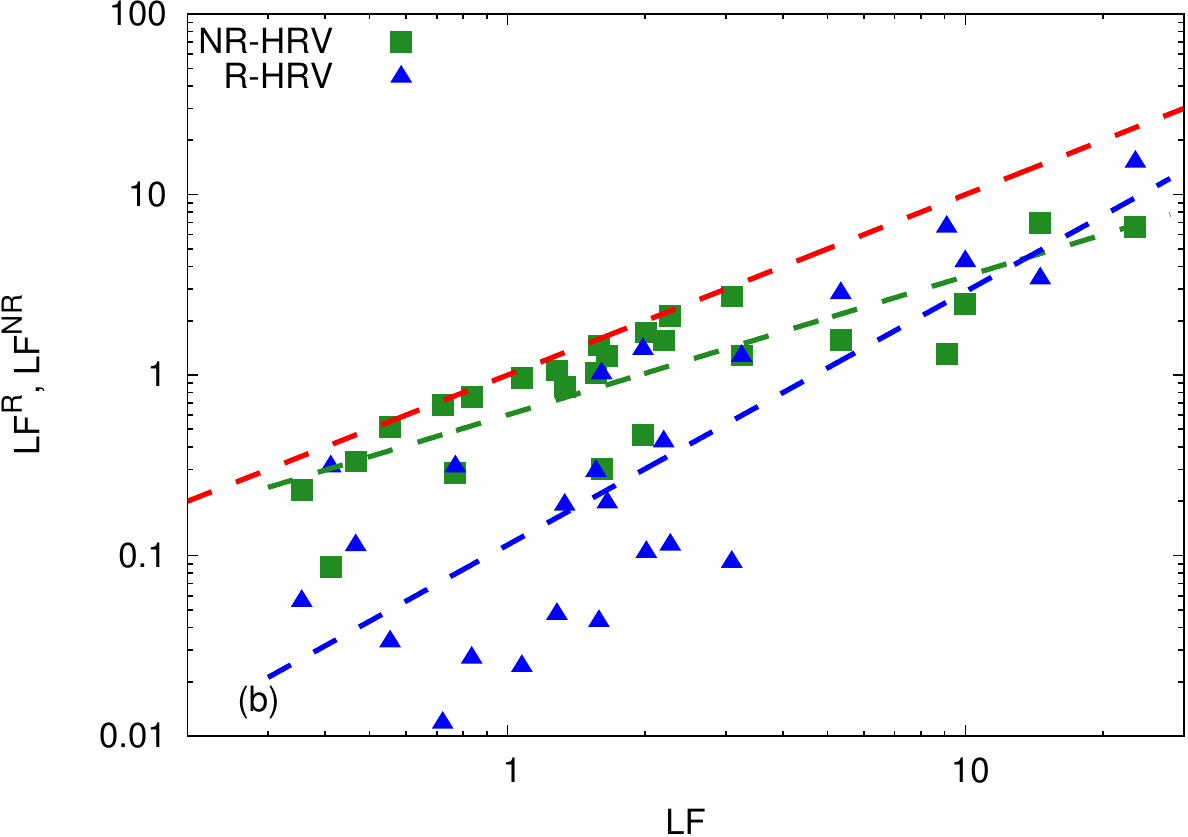}
\includegraphics[width=0.48\textwidth]{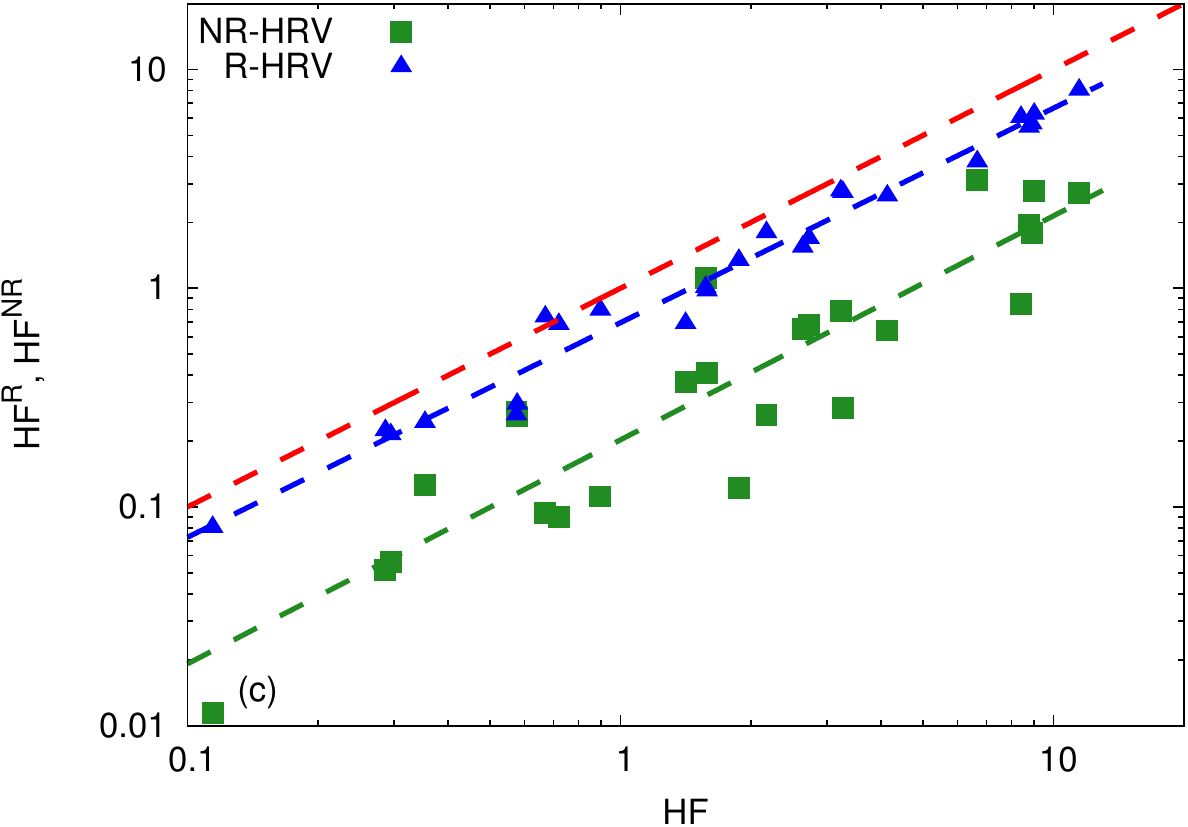}
\caption{Spectral power in very-low-frequency (a), low-frequency (b), 
and high-frequency (b) spectral ranges. 
Here the  R-HRV and NR-HRV components are shown vs the value of the corresponding 
power in the original time series. The red dashed line shows the diagonal, i.e. 
the values for the original time series). Other dashed lines show 
the power law fits (the correlation coefficients
for the fits are given in square brackets): 
$\text{VLF}^{\text{NR-HRV}}\approx 0.93\cdot (\text{VLF})^{1.01}$ [0.97]; 
$\text{VLF}^{\text{R-HRV}}\approx 0.016\cdot (\text{VLF})^{0.92}$ [0.58];  
$\text{LF}^{\text{NR-HRV}}\approx 0.6\cdot (\text{LF})^{0.77}$ [0.84]; 
$\text{LF}^{\text{R-HRV}}\approx 0.12\cdot (\text{LF})^{1.40}$ [0.79];  
$\text{HF}^{\text{NR-HRV}}\approx 0.20\cdot (\text{HF})^{1.02}$ [0.91]; 
$\text{HF}^{\text{R-HRV}}\approx 0.70\cdot (\text{HF})^{0.98}$ [0.99].}
\label{fig:fd}
\end{figure}

The frequency-domain measures are plotted in Fig.~\ref{fig:fd}.
Here we show the power in three frequency bands for R-HRV and NR-HRV 
series versus the corresponding powers of the original HRV 
(cf. dashed line which is the diagonal in the log-log representation). 
The VLF component (panel (a)) is strongly 
reduced in the R-HRV, while it is nearly the same as the original one in the NR-HRV. 
This is a clear indication that the very low-frequency variability on time scales 
larger than 20 s is not due to the respiration, but is caused 
by other physiological and external influences. Noteworthy, the results
for the VLF component are not very reliable due to a shortness
of the time series in our measurements.


In the low-frequency range (roughly time scales from 25 to 6 s, panel (b)) 
the reduction in the R-HRV component is not so strong. 
In fact, for some subjects the R-HRV component is even higher than the NR-HRV
component. It is known from
literature that slow respiration largly enhances HRV at 6 breaths per minute
where other LF rhythms are met \cite{Russo2017}. 
The high-frequency range (panel (c), time scales from 6 to 2 s) includes 
a typical period of breathing. 
Therefore, here the situation is the opposite to the cases of lower 
frequencies: the respiration component is only slightly less
than the original one, while the NR-HRV component is stronger reduced.

\subsection{Complexity measures}

Finally, we present the results of computation of complexity measures ApEn and SampEn.
Both entropies show similar behavior: their values for both R-HRV and NR-HRV are smaller 
then the entropy values of the original HRV signals, what means that the complexity of 
the total HRV is larger than that of its components. 
In almost all cases the NR-HRV signal shows larger regularity than the R-HRV component.

\begin{figure}[!hbt]
\centering
\includegraphics[width=0.48\textwidth]{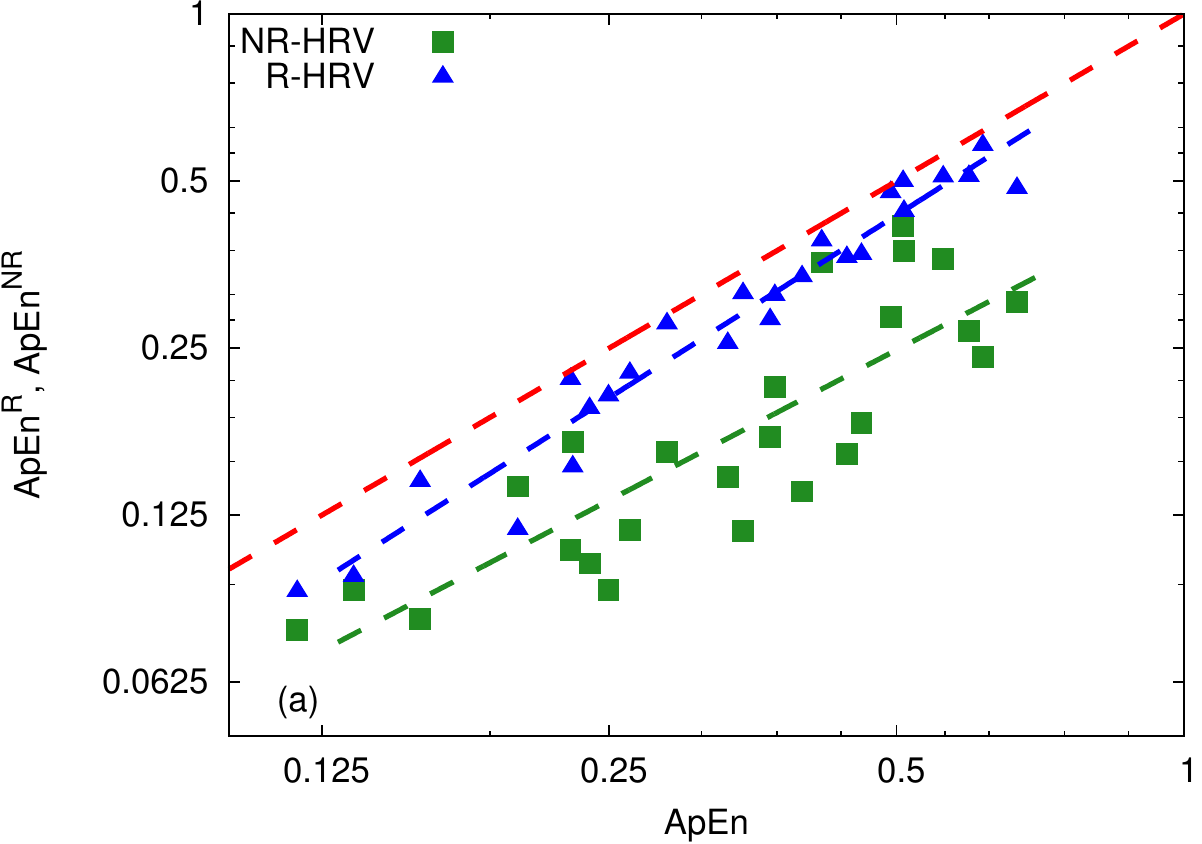}\hfill
\includegraphics[width=0.48\textwidth]{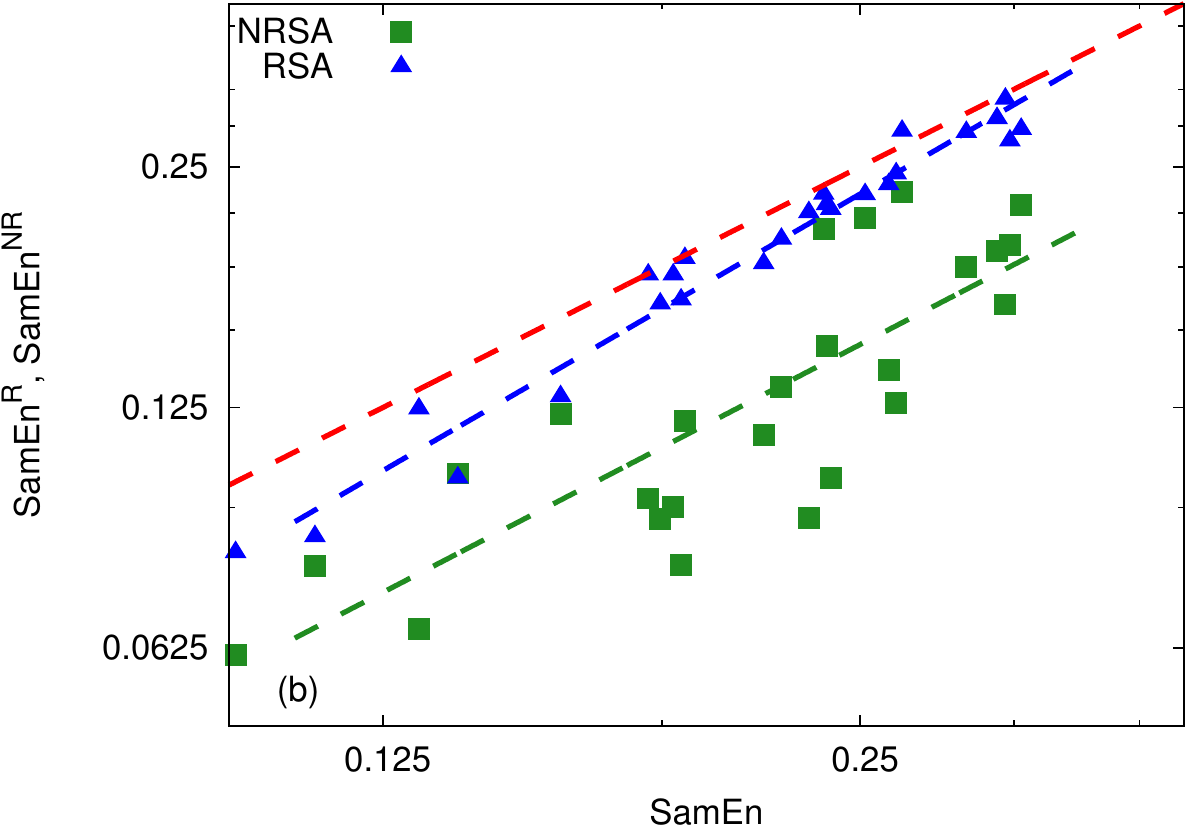}
\caption{Values of the approximate entropy (a) and of the sampled entropy (b) 
for the R-HRV and NR-HRV components, vs
the values of these quantities in the original time series (red dashed line 
is the diagonal). 
Other dashed lines show the power law fits (the correlation coefficients
for the fits are given in square brackets): 
$\text{ApEn}^{R}\sim \left(\text{ApEn}\right)^{1.09}$ [0.98];  
$\text{ApEn}^{NR}\sim \left(\text{ApEn}\right)^{0.9}$ [0.84]; 
$\text{SampEn}^{R}\sim \left(\text{SampEn}\right)^{1.15}$ [0.98];  
$\text{SampEn}^{NR}\sim \left(\text{SampEn}\right)^{1.03}$ [0.82]. 
One can see that for originally more complex signals (large 
values of SampEn and ApEn), reduction of complexity in the R-HRV component 
is smaller (the corresponding powers 
are larger than 1), but reduction in complexity
in the NR-HRV component is larger (the powers are closer to 1).}
\label{fig:en}
\end{figure}

\section{Conclusions and discussion}
We have presented and illustrated a nonlinear technique which enables
disentanglement of the RR-intervals series into the respiratory-related 
component, R-HRV, and the rest, NR-HRV. 
The procedure can be performed if simultaneous 
measurements of an ECG and of a respiratory signal are available. 
Our method is based on the coupled oscillators model, and thus is 
inherently nonlinear. In particular, this means that our procedure is 
not a simple decomposition (in the sense of signal decomposition techniques), 
i.e. $\text{R-HRV}+\text{NR-HRV}\ne \text{HRV}$. 
The spectral analysis confirms methodical validity of our approach, 
demonstrating that the R-HRV component correctly describes peaks in the 
power spectra at the respiratory-related frequencies. 
We suggest to use this approach as a 
universal preprocessing technique which allows a researcher to concentrate 
on particular properties of the HRV data. Moreover, the technique can be
used for investigation of other physiological rhythms if bivariate or 
multivariate measurements are available.

The results presented in Figs.~\ref{fig:rmssd}-\ref{fig:en} show
that some HRV measures appear to be dominated by one of the components. 
So, Figs.~\ref{fig:rmssd}b,c show that the values of $\text{logRSA}^{\text{R}}$ 
and of $\text{pNN50}^{\text{R}}$ are 
very close to $\text{logRSA}$ and $\text{pNN50}$, respectively.
(Fig.~\ref{fig:fd})a. 
In particular, this indicates that $\text{logRSA}$ and $\text{pNN50}$ 
may be more efficient for quantification of the respiratory-related 
component than other measures, if raw data are used. However, $\text{pNN50}$
suffers from a saturation effect in subjects where HRV is generally low.
However, verification of this hypothesis requires
further analysis with data from different groups of subjects.
Moreover, Figs.~\ref{fig:rmssd}-\ref{fig:en} clearly demonstrate that  
generally computation of standard HRV measures from original 
data and its disentangled components yield different results, 
and this difference can be essential. 
 
We notice that the technique requires rather high-quality ECG records and 
quite a demanding preprocessing, related to the phase estimation. 
Therefore we consider the current results as a ``proof of principle''.
A very useful practical improvement, now in progress, would be development of 
an approximate disentanglement algorithm that can be performed on the basis 
of RR-series only. 
Another essential potential development would be to incorporate into the model 
the amplitude variations of the respiratory signal, 
modifying Eq.~(\ref{eq:gpd7}) to
\begin{equation}
\dot\vp=\w+ A(t)Q(\varphi,\psi)+ \xi(\vp,t)\;,
\label{eq:gpd7mod}
\end{equation}
where $A(t)$ is the instantaneous amplitude that can be easily extracted by 
means of the Hilbert transform.
However, for this purpose another type of measurements is required, where
the amplitude of the measured respiratory signal can be attributed to true 
amplitude of the respiratory force affecting the cardio-vascular system. 
  
The new method promises an improved determination of vagal tone for medicine 
and prevention, based on elaborated mathematical decomposition of available data.
It also allows for a better understanding of the non-respiratory components of 
HRV in general. The parasympathetic nervous system is an important part of our control 
center for silent inflammation (also coined ``secret killer'' by Time 
magazine~\cite{Gorman2004}).  
Cellular receptors allowing communication
of autonomic nervous system and immune cells have been found in the past
decade and vagal activity has been proven to control immune activity.
Therefore, a noninvasive quantification of vagal tone will become important
in several fields of medicine in the near future for diagnostic as well as
therapeutic purposes. The method described in this paper might be a
valuable contribution for such a highly desired accurate measurement tool. 
The amount of vagal tone describes the 
ability of the organism to recover after inflammatory states, 
condition very important not only for our 
well-being and healthy aging, but for the prevention of serious chronic 
diseases like atherosclerosis, neurodegenerative diseases or cancer. 
Application of the method to medical diagnostics requires, however, 
future measurements from the corresponding groups of patients.

\section{Acknowledgments}
\c{C}T was financially supported by the European Union’s Horizon
2020 research and innovation programme under the Marie
Sklodowska-Curie Grant Agreement No. 642563 (COSMOS).
Development of methods presented in Section 2 was supported by
the Russian Science Foundation under Grant No. 17-12-01534.
We acknowledge helpful discussion with Prof. J.~Schaefer.

\newpage

\end{document}